\documentclass[12pt]{article}
\usepackage{latexsym}
\usepackage{graphics}
\usepackage{epsfig}

\newcommand{\bmat}{\left(\begin{array}}
\newcommand{\emat}{\end{array}\right)}
\newcommand{\beq}{\begin{equation}}
\newcommand{\eeq}{\end{equation}}

\newcommand{\drawsquare}[2]{\hbox{%
\rule{#2pt}{#1pt}\hskip-#2pt
\rule{#1pt}{#2pt}\hskip-#1pt
\rule[#1pt]{#1pt}{#2pt}}\rule[#1pt]{#2pt}{#2pt}\hskip-#2pt
\rule{#2pt}{#1pt}}

\newcommand{\fund}{\raisebox{-.5pt}{\drawsquare{6.5}{0.4}}}
\newcommand{\Ysymm}{\raisebox{-.5pt}{\drawsquare{6.5}{0.4}}\hskip-0.4pt%
        \raisebox{-.5pt}{\drawsquare{6.5}{0.4}}}
\newcommand{\Yasymm}{\raisebox{-3.5pt}{\drawsquare{6.5}{0.4}}\hskip-6.9pt%
        \raisebox{3pt}{\drawsquare{6.5}{0.4}}}
\newcommand{\antifund}{\overline{\fund}}

\def\yzero{\smash{\hbox{$y\kern-4pt\raise1pt\hbox{${}^\circ$}$}}}

\def\-{\hphantom{-}}
\def\ov{\overline}
\def\s2{\frac{1}{\sqrt2}}

\def\beq{\begin{equation}}
\def\eeq{\end{equation}}
\def\beqa{\begin{eqnarray}}
\def\eeqa{\end{eqnarray}}

\def\IF{\relax{\rm I\kern-.18em F}}
\def\II{\relax{\rm I\kern-.18em I}}
\def\IP{\relax{\rm I\kern-.18em P}}

\def\Dsl{\,\raise.15ex\hbox{/}\mkern-13.5mu D} 

\def\IC{\bf C}
\def\IZ{\bf Z}

\def\z2z2{$\IC^3/(\IZ_2\times\IZ_2)$}

\def\o{\omega}

\def\s{\sigma}

\def\z{\zeta}

\def\bo{{\raise-.3ex\hbox{\large$\Box$}}}               
\def\face{{\raise.2ex\hbox{$\displaystyle \bigodot$}\mskip-2.2mu \llap {$\ddot
        \smile$}}}                                      

\def\leftrightarrowfill{$\mathsurround=0pt \mathord\leftarrow \mkern-6mu
        \cleaders\hbox{$\mkern-2mu \mathord- \mkern-2mu$}\hfill
        \mkern-6mu \mathord\rightarrow$}       
\def\dvec#1{\vbox{\ialign{##\crcr
        \leftrightarrowfill\crcr\noalign{\kern-1pt\nointerlineskip}
        $\hfil\displaystyle{#1}\hfil$\crcr}}}           

\def\beq{\begin{equation}}
\def\eeq{\end{equation}}

\def\beqx{\begin{displaymath}}
\def\eeqx{\end{displaymath}}

\def\beqa{\begin{eqnarray}}
\def\eeqa{\end{eqnarray}}

\begin{document}

\DeclareGraphicsExtensions{.jpg,.pdf,.mps,.png}
\begin{flushright}
\baselineskip=22pt
UPR-1159-T, hep-th/0607238 \\
\end{flushright}

\begin{center}
\vglue 0.5cm

{\large\bf New  Grand Unified Models with Intersecting D6-branes, Neutrino
Masses, and Flipped SU(5)} \vglue 1cm { Mirjam Cveti\v c$^a$ and Paul
Langacker$^{a,b}$} \vglue 1cm { $^a$ Department of Physics and Astronomy,
University of Pennsylvania, \\Philadelphia, PA 19104-6396, USA \\

$^b$ School of Natural Sciences, Institute for Advanced Study,  \\
             Einstein Drive, Princeton, NJ 08540, USA\\}
\end{center}

\thispagestyle{empty}


\begin{abstract}
{
\scriptsize
We construct new supersymmetric  $SU(5)$ Grand Unified Models  based on ${\bf
Z_4\times Z_2}$
 orientifolds with intersecting D6-branes. Unlike constructions based on ${\bf Z_2\times Z_2}$
  orientifolds, the orbifold images of the three-cycles wrapped by D6-branes correspond to new configurations
  and thus allow for models  in which,  in addition to the chiral sector
  in ${\bf 10}$ and
 ${\bf {\bar 5}}$  representations of $SU(5)$, only,  there can be new sectors
with  $\left({\bf 15}\ +\ {\bf{\overline{ 15}}}\right)$
  and $\left({\bf 10}\ +\ {\overline{\bf10}}\right)$ vector-pairs.
  We  construct an example of  such a globally consistent, supersymmetric  model with   four-families,
  two  Standard Model Higgs pair-candidates and the  gauge symmetry $U(5)\times U(1)\times
  Sp(4)$.
  In a $N=2$ sector, there are  $5\times\left({\bf 15}\ +\ {\bf{\overline{15}}}\right)$  and
   $1\times \left({\bf 10}\ +\ {\bf{\overline{10}}}\right)$ vector pairs, while another $N=1$ sector contains
   one vector-pair of ${\bf 15}$-plets.
     The  $N=2$ vector pairs  can obtain a   large mass dynamically by parallel D6-brane
     splitting in a particular two-torus.  The ${\bf 15}$-vector-pairs provide, after symmetry breaking
    to the Standard Model (via parallel D-brane splitting),
   triplet pair candidates which  can in  principle  play a role in  generating
     Majorana-type  masses for left-handed neutrinos, though the necessary
     Yukawa couplings are absent in the specific construction.
This model can also be interpreted as a flipped $SU(5)\times U(1)_X$ Grand
Unified Model where the ${\bf 10}$-vector-pairs can play the role of Higgs
fields, though again there are phenomenological difficulties for the specific
construction.
}
\end{abstract}

\newpage

\section{Introduction}
The explanation of the  origin of small neutrino masses in string
constructions is a notoriously difficult problem.
 In particular, most of the
intersecting D brane constructions of the semi-realistic Standard Model string
vacua allow for Dirac neutrino masses; however, it is typically difficult to
ensure small Dirac neutrino masses while on the other hand providing for an
acceptable  mass hierarchy in the quark and charged lepton sector
\cite{Ibanez:2001nd, CLS,Antoniadis:2002qm, Kokorelis,Dutta}. No examples of
intersecting D brane constructions leading to Majorana masses have been given.
Within heterotic string theory, it is possible in principle to realize the
usual minimal seesaw model\footnote{For reviews of neutrino mass mechanisms,
see, for example,
 \cite{Mohapatra:2005wg,lep}.}. In practice, however, it is difficult to simultaneously
 generate a large Majorana mass for the singlet neutrino and a Dirac mass coupling for the
 doublet and singlet neutrinos, while preserving supersymmetry at large scales and respecting
 the necessary consistency conditions for the string construction \cite{Font:1989aj}-\cite{gkln},
 with the few examples being non-minimal (i.e., involving a higher power of the heavy mass in the
 denominator \cite{CL92,Coriano:2003ui}) or not GUT-like \cite{Ellis:1997ni,Ellis:1998nk,Kim:2004pe},
 and often invoking additional dynamical assumptions. A systematic survey of a class
 of $\bf Z_3$ orbifold constructions did not find a single example of a minimal seesaw \cite{gkln},
 and a study of $\bf Z_6$ constructions did not find any examples to the order considered if $R$-parity
 is imposed \cite{Kobayashi:2004ya}. Similar problems may occur for theories with additional
 low energy symmetries \cite{KLL}.

Within the framework of particle physics model building,  one intriguing
possibility of generating small Majorana masses  is  via  vector pairs in
$\left( {\bf 3}\ + \ {\bf \overline3}\right)$ representations of $SU(2)_L$
with unit hypercharge  \cite{Ma:1998dx,Hambye:2000ui,Rossi:2002zb}.
If the ${\bf 3}$ couples to a pair of lepton doublets and
the $\bf \overline3$ to a pair of up-type Higgs doublets (or the ${\bf 3}$ to a pair of
down-type Higgs), then lepton number is violated. If there is also a large
supersymmetric mass $M_T$ for the ${\bf 3}\ + \ {\bf \overline3}$ pair, then the
neutral component of the ${\bf 3}$ will acquire a tiny expectation value of order
$|\langle H_u^0\rangle|^2/M_T$, leading to the so-called type II seesaw mechanism.
 (If there is no $ {\bf \overline3 \ H_uH_u}$ coupling, the ${\bf 3\ H_dH_d}$
coupling generates a mass of order $M_T^{-2}$ \cite{KLL}.) The possibility of
realizing such a triplet seesaw mechanism within heterotic string
constructions was considered in \cite{bnelson}. An $SU(2)_L$ triplet with unit
hypercharge could be obtained by a diagonal embedding of $SU(2)_L$ into
$SU(2)\times SU(2)$ (i.e., a higher affine Kac-Moody level construction). It
was shown that  such a construction would most naturally lead to an
off-diagonal mass matrix, and therefore to distinct phenomenological features
(e.g., an inverted hierarchy with observable neutrino-less double beta decay
and a mixing $U_{e3}$ close to the present experimental limit), very different
from
 triplet models motivated from bottom-up or non-string motivations.
 Explicit constructions were given with many, but not all, of the necessary ingredients.

Higgs triplet pairs with unit hypercharge  can  arise as a decomposition of
 $\left({\bf 15}\ +\ {\bf{\overline{15}}}\right)$ pairs of the $SU(5)$ Grand
Unified Theory (GUT) \cite{Georgi}. (For reviews see \cite{Langacker,Nath}.)
The purpose of this paper is to realize this mechanism within explicit,
globally consistent supersymmetric string constructions. The concrete
realization  is based on  intersecting D6-brane constructions on toroidal
orbifolds.  (For a review see \cite{BCLS} and references therein.) This
framework provides a natural mechanism to realize supersymmetric $SU(5)$ GUT
constructions \cite{CSUII,CPS}.\footnote{For the original work on
non-supersymmetric chiral intersecting D-branes, see
\cite{{BlumI},{AldazabalI}, {AldazabalII},{BlumII}}. For chiral supersymmetric
ones, see \cite{{CSUI}, {CSUII}} and also \cite{Angelantonj}. For
supersymmetric chiral constructions within Type II rational conformal field
theories, see \cite{Dijkstra,Kiritsis}  and references therein. For the study
of  the landscape of intersecting D-brane constructions, see \cite{BlI,DT}.

For flipped $SU(5)$ GUT constructions, see \cite{NanopoulosI,NanopoulosII}.
For recent  GUT constructions with intersecting D6-branes, see also \cite{FK}
and references therein. For related studies of features of GUT's in the Type
II context, see \cite{Tatar,Berenstein}. For proton decay calculations within
intersecting D6-brane constructions (and their strong coupling limits), see
\cite{KW,CR}.} In these constructions the ${\bf 10}$-plets (and ${\bf
15}$-plets) arise from the intersections of the $U(5)$ D6-brane configuration
and its orientifold image. The appearance of ${\bf 15}$-plets turns out to be
ubiquitous  in such constructions \cite{CPS}. The major drawback of these
constructions is the absence of the  up-quark Yukawa couplings  to the
Standard Model  (SM) Higgs; they are zero in perturbative Type IIA theory
\cite{CPS,Berenstein},  due the conservation of the ``anomalous'' $U(1)$ part
of the $U(5)$  GUT symmetry.

The known  supersymmetric  GUT constructions with intersecting D6-branes are
based on ${\bf Z_2\times Z_2}$ orientifolds, where the orbifold images of
three-cycles, which  are ``inherited'' from the toroidal ones, are the same as
the original three-cycles.  Therefore,  D6-brane configurations that wrap such
three-cycles result in  massless open-string sectors that effectively arise
from a single set of  D6-brane configurations, inherited from the toroidal
ones. The massless spectrum in each such sector  is either associated with the
$N=1$ supersymmetric chiral sector or purely non-chiral $N=2$ supersymmetric
ones. As a consequence, intersecting D6-brane constructions on ${\bf Z_2\times
Z_2}$ orientifolds  cannot account for the appearance of $N=2$ vector pairs of
${\bf 15}$-plets (and/or  $N=2$ vector pairs of ${\bf 10}$-plets), without the
introduction of the chiral ``excess'' of  the same number of  ${\overline{\bf
15}}$ as there are chiral ${\bf 10}$-plets.   Namely, the ${\bf 10}$-plets
should be chiral to be identified with the fermion families. However, since
they arise from the same sector as ${\bf 15}$'s, the latter are also
necessarily chiral. This feature also  applies to the flipped $SU(5)$
constructions \cite{NanopoulosI,NanopoulosII}, which require in addition to
chiral matter in $\bf 10$ and ${\bf\bar 5}$ representations of $SU(5)$, also
additional GUT Higgs multiplets in $\left( {\bf 10}\ \ +\ \overline{{\bf
10}}\right)$ representations. However,  within ${\bf Z_2\times Z_2}$
orientifold constructions the appearance of the GUT Higgs pairs necessarily
requires a net number of chiral ${\overline{\bf 15}}$'s, which is the same as
the number of chiral ${\bf 10}$-plets, and thus the $SU(5)$ anomaly
cancellation requires additional ${\bf 5}$'s.

In this paper we therefore turn to constructions of supersymmetric $SU(5)$
GUT's that are based on orientifolds whose  orbifold action produces new
D6-brane configurations, and thus in addition to the original brane
configurations, with say, a chiral sector, one now has new sectors, associated
with the orbifold images, that can provide, say, non-chiral sectors. In order
to demonstrate the existence of such constructions, we shall focus on a
specific orientifold, which we choose for simplicity to be the ${\bf Z_4\times
Z_2}$ orientifold. In addition, we choose only  the three-cycles inherited
from the toroidal constructions, i.e., for simplicity, we do not include
fractional D-brane configurations, associated with the three-cycles wrapping
orbifold singularities. A class of such orientifold constructions  was
discussed in detail in \cite{Honecker}, with a goal to obtain three-family
Standard Models. Here, our aim is to employ such an orientifold to construct
supersymmetric GUT models with the features described above. In particular, we
shall describe in detail an explicit, globally consistent supersymmetric
construction with four-families and the gauge group structure of $U(5)\times
U(1) \times Sp(4)$. Note that this explicit construction is meant to
demonstrate specific features GUT spectrum, that in particular the non-chiral
sector  allows for matter candidates that may have interesting implications
for neutrino masses.
 [Within  the framework of  flipped $SU(5)$ GUT constructions we shall see that
the construction of the type presented in this paper can also be interpreted
as a flipped $SU(5)$ GUT   with  {\it no} net ${\bf 15}$'s, while the GUT
Higgs candidates of flipped $SU(5)$, i.e., $({\bf 10}+\overline{\bf
10})$-pairs, arise from the $N=2$ sector of the construction.] Of course, one
can also pursue constructions on other orbifolds, such as, e.g.,
\cite{Blumenhagen,HO}, and involve there more general cycles with fractional
D6-branes, e.g., \cite{Blumenhagen,HO,BCMS}, which is a topic of further
research.

The paper is organized as follows. In Section   \ref{z2z4orien} we discuss
features of the ${\bf Z_4\times Z_2}$ orientifold, such as orbifold and
orientifold actions and the corresponding ${\rm O}6$-planes. In Section
\ref{tools} we discuss in detail the spectrum, global consistency conditions
and supersymmetry conditions for the open string sector of D6-branes wrapping
the three-cycles of the ${\bf Z_N\times Z_M}$ orientifold, emphasizing the
geometric aspects of the spectrum and consistency conditions for three-cycles
inherited from the six-torus. This section also serves as a set-up for
intersecting D6-brane constructions on three-cycles inherited from  the
six-torus for more general orientifolds than the one discussed in this paper.
In Section \ref{GUT} we discuss general features of the spectrum and couplings
of the GUT models  in the intersecting D6-brane constructions. In Section
\ref{modelc} we provide explicit expressions for the global consistency,
K-theory constraints and supersymmetry conditions, as well as intersection
numbers of the massless matter supermultiplets  for open string sectors of a
${\bf Z_4\times Z_2}$ orientifold. In Section \ref{explm} we construct an
explicit example of a supersymmetric, globally consistent four-family GUT
model and discuss in detail the open string sector massless spectrum as well
as features of Yukawa couplings. In Subsection \ref{flip} we also address the
interpretation of the spectrum in the context of flipped $SU(5)xU(1)_X$ and
show that the choice of the $U(1)_X$ gauge symmetry  is non-anomalous with the
massless  gauge boson. In Section \ref{conc} we summarize the results at the
spectrum level and point toward future constructions that may overcome
phenomenological difficulties at the level of Yukawa couplings.

\section{${\bf Z_4\times Z_2}$ Orientifold}\label{z2z4orien}

The construction is based on the
  ${\bf T}^6 /({\bf Z_4\times Z_2})$
orientifold.
 We
consider ${\bf T}^{6}$ to be a six-torus factorized as ${\bf T}^{6} = {\bf
T}^{2} \times{\bf T}^{2} \times {\bf T}^{2}$ whose complex coordinates are
$z_i$, $i=1,\; 2,\; 3$ for the $i$-th two-torus, respectively. The $\theta$
and $\omega$ generators for the orbifold group $\bf Z_{4}\times Z_2$  are
 associated with  twist vectors $(1/4,-1/4,0)$ and
$(0,1/2,-1/2)$, respectively; they act  on the complex coordinates of
${\bf T}^6$ as
\begin{eqnarray}
& \theta: & (z_1,z_2,z_3) \to ({\rm i}\, z_1,-{\rm i}\, z_2,z_3)~,~ \nonumber \\
& \omega: & (z_1,z_2,z_3) \to (z_1,-z_2,-z_3)~.~\, \label{orbifold}
\end{eqnarray}
The orientifold projection is implemented by gauging the symmetry $\Omega R$,
where $\Omega$ is world-sheet parity, and $R$ acts as
\begin{eqnarray}
 R: (z_1,z_2,z_3) \to ({\ov z}_1,{\ov z}_2,{\ov
z}_3)~.~\, \label{orientifold}
\end{eqnarray}

\renewcommand{\arraystretch}{1.4}
\begin{table}[t]
\caption{Wrapping numbers of the four O6-planes, fixed under
 the ${\bf Z_2\times Z_2}$ action of $\theta^2$  and  $\omega$ generators. $b$ is equal to 0 and
$\frac{1}{2}$ for rectangular and tilted third two-torus, respectively. }
\vspace{0.4cm}
\begin{center}
\begin{tabular}{|c|c|c|}
\hline
  Orientifold Action & O6-Plane & $(n^1,m^1)\times (n^2,m^2)\times
(n^3,m^3)$\\
\hline
    $\Omega R$& 1 & $(1,0)\times (1,1)\times
(4^b,-2b)$ \\
\hline
    $\Omega R\omega$& 2& $(1,0)\times (1,-1)\times
(0,1)$ \\
\hline
    $\Omega R\theta^2\omega$& 3 & $(0,-1)\times
(1,1)\times
(0,1)$ \\
\hline
    $\Omega R\theta^2$& 4 & $(0,-1)\times (-1,1)\times
    (4^b,-2b)$ \\
\hline
\end{tabular}
\end{center}
\label{table1O6}
\end{table}

\begin{figure}
\begin{center}
\scalebox{0.55}{{\includegraphics{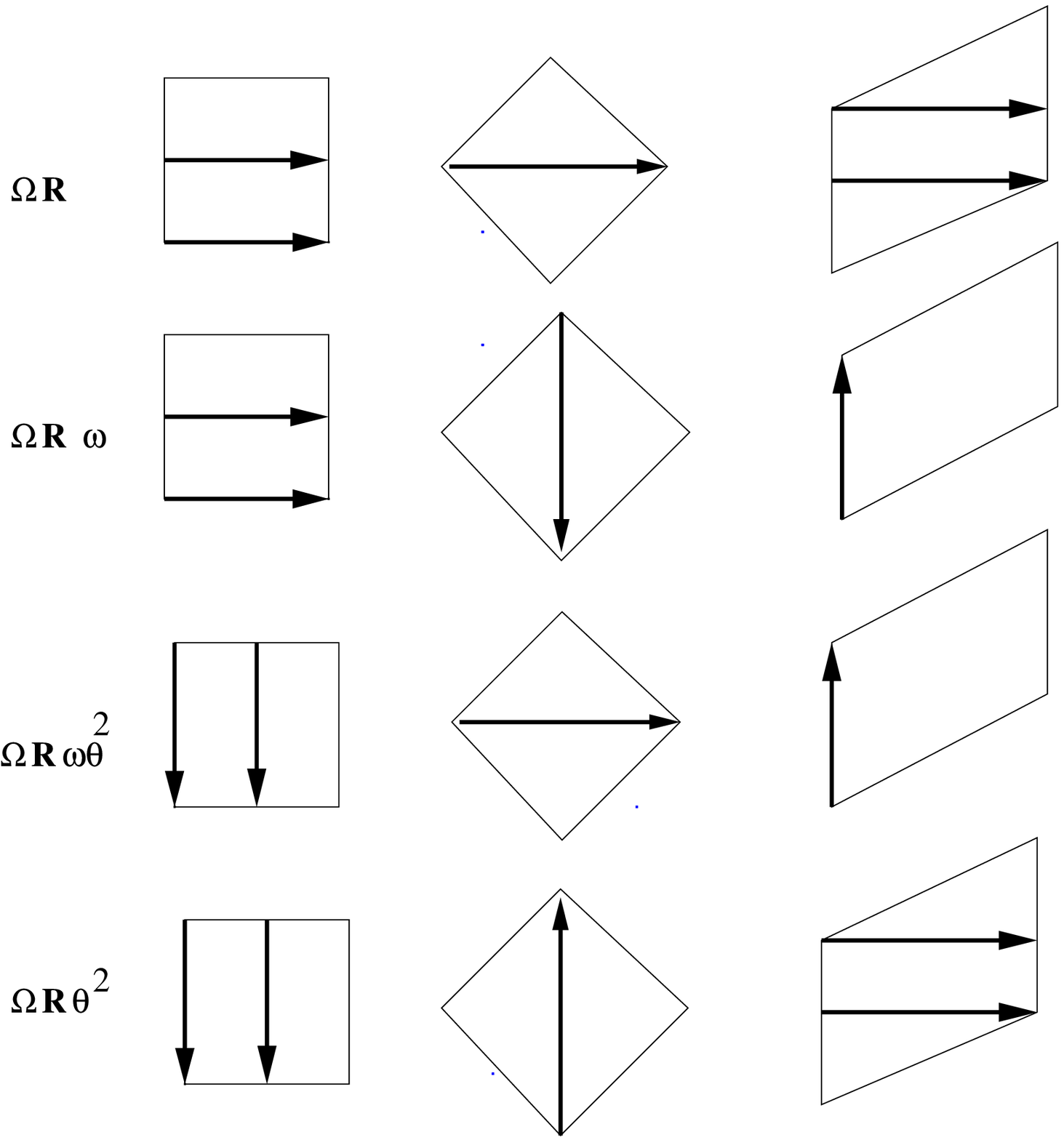}}}
\end{center}
\caption[]{\small The locations of O6-planes,  fixed under the orientifold
actions $\Omega R$,  $\Omega R \omega$, $\Omega R
\omega \theta^2$, and $\Omega R \theta^2$
 (denoted by  bold solid lines)
for a  factorized six-torus with  the third two-torus tilted. }

\label{MajorO1}
\end{figure}

\renewcommand{\arraystretch}{1.4}
\begin{table}[t]
\caption{Wrapping numbers of the  $\theta$ images of the four O6-planes, fixed
under the action of $\theta$ and $\omega$. $b$ is equal to $0$ and
$\frac{1}{2}$ for rectangular and tilted third two-torus, respectively.}
\vspace{0.4cm}
\begin{center}
\begin{tabular}{|c|c|c|}
\hline
  Orientifold Action & $\theta$ {\rm O}6-Plane & $(n^1,m^1)\times (n^2,m^2)\times
(n^3,m^3)$\\
\hline
    $\Omega R\theta$& 1 & $(1,-1)\times (0,1)\times
(4^b,-2b)$ \\
\hline
    $\Omega R\omega\theta$& 2& $(1,-1)\times (1,0)\times
(0,1)$ \\
\hline
    $\Omega R\theta^3\omega$& 3 & $(-1,-1)\times
(0,1)\times
(0,1)$ \\
\hline
    $\Omega R\theta^3$& 4 & $(1,1)\times (1,0)\times
    (4^b,-2b)$ \\
\hline
\end{tabular}
\end{center}
\label{tableth1}
\end{table}

\begin{figure}
\begin{center}
\scalebox{0.55}{{\includegraphics{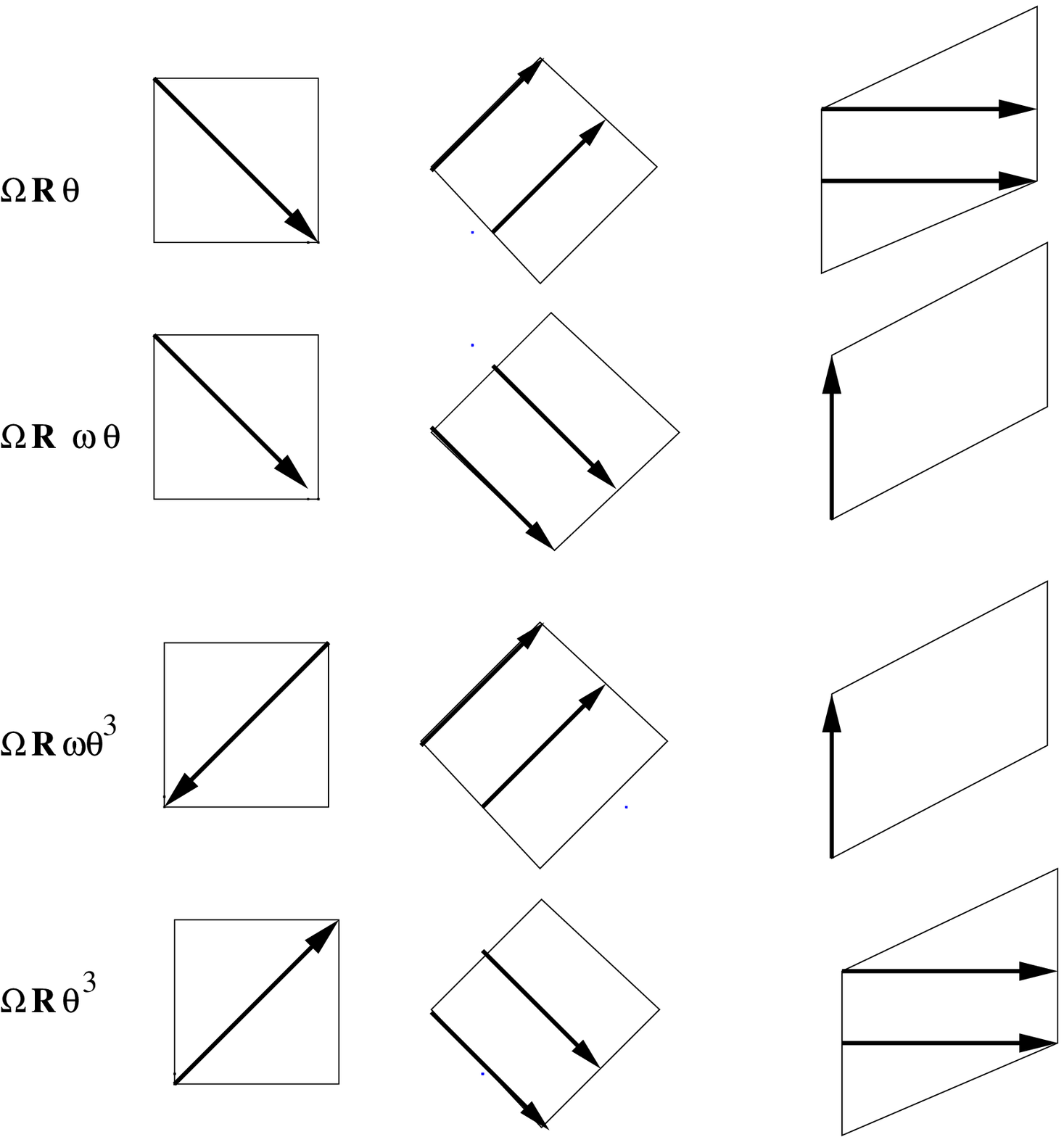}}}
\end{center}
\caption[]{\small The locations of O6-planes fixed under the orientifold
actions $\Omega R\theta$,  $\Omega R \omega\theta$,
$\Omega R \omega \theta^3$, and $\Omega R \theta^3$
 (denoted by  bold solid lines)
for the factorized  six-torus with the third two-torus tilted.}
\label{Majorth1}
\end{figure}

 We briefly
review  the basics of the constructions first for the  model
\cite{Honecker,Honeckerp} where the torus configurations and the corresponding
$O6$ planes as well as their images are depicted in Figures \ref{MajorO1} and
\ref{Majorth1}, respectively. [One can  also consider a somewhat different
construction  with the second two-torus modified to be of the same type as the
first two-torus and vice versa. In addition,  constructions on other type of
toroidal orbifolds, such as ${\bf Z_4}$-orientifold  \cite{Blumenhagen}, ${\bf
Z_6}$-orientifold \cite{HO} and ${\bf Z_2\times Z_2}$-orientifold with torsion
\cite{BCMS},  and inclusion of  more general cycles at orbifold singularities
\cite{Blumenhagen,HO,BCMS}, resulting in fractional D-brane configurations
branes are of interest. These aspects of constructions will be discussed
elsewhere.]

When a specific brane configuration is invariant under
these orbifold actions, the corresponding Chan-Paton factors are subject to
their  projections, as discussed in the following subsections.
 [The fact that  D6-branes are invariant under orbifold
projections does not imply  that their intersection points will be.
 The final spectrum, however, turns out to be rather insensitive
to this subtlety in the case of the ${\bf T}^6 /(\IZ_2 \times \IZ_2)$
orientifold construction. See  \cite{CSUII} for further discussions.]

There are four kinds of orientifold 6-planes (O6-planes) due to the action of
$\Omega R$, $\Omega R\omega$, $\Omega
R\theta^2\omega$, and  $\Omega R \theta^2$, respectively. Their configurations are tabulated
in
Table \ref{table1O6}  and presented geometrically  in Figure  \ref{MajorO1},
respectively.  The corresponding images under the  orbifold actions $\theta$
are given in Table \ref{tableth1}  and Figure  \ref{Majorth1}.

\section{Tools for ${\bf Z_N\times Z_M}$
orientifolds}\label{tools}

\subsection{Massless Open String Spectrum} \label{ssms}

For the orientifold models with intersecting D6-branes  wrapping three-cycles,
inherited from the six-torus, the chiral spectrum, arising from open string
sectors, can be determined  geometrically  from the intersection numbers of
the three-cycles the D6-branes are wrapped around.  For $N_a$ D6-branes that
wrap three-cycles, not invariant under the anti-holomorphic involution, the
gauge symmetry  is $ U(N_a)$. For this case the general rule for determining
the massless left-handed chiral spectrum is presented in Table \ref{tcs} (for
details see, e.g., \cite{BCLS}).
\begin{table}
\caption{Chiral spectrum for intersecting D6-branes wrapping three cycles
$\pi_a$ \cite{BCLS}. We choose a convention that the negative intersecting
numbers below correspond to the left-handed chiral superfields in  the
representations displayed in the first column.} \centering \vspace{3mm}
\label{tcs}
\begin{tabular}{|c|c|}
\hline
Representation  & Multiplicity \\
\hline $\Yasymm_a$
 & ${1\over 2}\left(\pi_a\circ \pi_{a'}+
\pi_a \circ\pi_{{\rm O}6}\right)$  \\
$\Ysymm_a$
     & ${1\over 2}\left(\pi_{a}\circ \pi_{a'}- \pi_a\circ \pi_{{\rm O}6}\right)$   \\
$(\fund_a,\antifund_b)$
& $\pi_a\circ \pi_{b}$   \\
 $(\fund_a,\fund_b)$
 & $\pi_a\circ \pi_{b'}$
\\
\hline
\end{tabular}
\end{table}
Open strings stretched between a D-brane and  its $\o\sigma$ image are the
only ones left invariant under the combined operation $\Omega\o\sigma
(-1)^{F_L}$. Here $F_L$ refers to the left-moving world-sheet fermion number.
Therefore, they transform in the antisymmetric or symmetric representation of
the gauge group, indicating more general representations in an orientifold
background. These representations play an important role in the construction
of  $SU(5)$ Grand Unified Models (GUT's).

To apply Table \ref{tcs}  to concrete models, one has to compute the
intersection numbers of three-cycles. We  focus only on the three-cycles
$\pi_a$ that are ``inherited'' from the three-cycles of the six-torus. In the
case of toroidal   orbifolds, such as  $T^6/(\bf{Z_N}\times \bf{Z_M})$, the
application  and geometric interpretation of the  Table \ref{tcs} for such
cycles can be made explicit.

The spectrum of Table \ref{tcs}  implies the computation for the intersection
numbers on the  orbifold, and not on the ambient  torus) (see also
\cite{BCLS}). For three-cycles $\pi_a$ on the orbifold space,  which are
inherited from the torus, the three-cycles $\pi^T_a$ on the torus are arranged
in orbits of length $N$ and $M$, under  the $\bf{Z_N}\times \bf{Z_M} $
orbifold group, i.e.,
\begin{equation} \pi_a = \sum_{i=0}^{N-1} \sum_{j=0}^{M-1}
\theta^i\,\omega^{j}\, \pi^T_a\ ,
\end{equation} where $\theta^i$ and $\omega^{j}$ denote the generators of $\bf{Z_N}$ and $\bf{Z_M}$,
respectively. The definition of the orientifold image cycle $\pi_{a'}$ is
analogous, with $\pi^T_a$  replaced by the orientifold image on the torus,
denoted by $\pi^T_{a'}$.  The three-cycles $\pi_{{\rm O}6}$ of the ${\rm O}6$
planes, fixed under the orientifold action, take the following analogous form:
\begin{equation} \label{oriencycle}  \pi_{{\rm O}6} = \sum_{i=0}^{N-1} \sum_{j=0}^{M-1}
\theta^i\,\omega^{j}\, \pi^T_{{\rm O} 6} \ .
\end{equation}
 Such orbits can then be considered as a three-cycle of the orbifold, where
the intersection number is given by
\begin{equation} \label{orcycle} \pi_a\circ\pi_b={1\over {N\, M}} \left(\sum_{i=0}^{N-1}\,\sum_{j=0}^{M-1}
\theta^i\,\omega^{j} \pi^T_a
  \right) \circ \left(\sum_{i'=0}^{N-1}\, \sum_{j'=0}^{M-1} \theta^{i'}\, \omega^{j'}\,
\pi^T_b \right)\ . \end{equation}
 [In addition to these untwisted
three-cycles, certain twisted sectors of the orbifold action can give rise to
so-called twisted three-cycles, associated with the fractional D-branes at
orbifold singularities, but we shall not include these cycles in our
consideration.]
Table \ref{tcs} only gives the chiral spectrum
of an intersecting D6-brane model. To compute the non-chiral spectrum one has
to employ the enhanced supersymmetry  associated with  a specific $T^2$, as
will be discussed in a concrete case for the ${\bf Z_4\times Z_2}$ orientifold in
Section   \ref{modelc}.

For a factorizable product of three-one cycles on the six-torus, $\pi^T_a$
 can be explicitly written in terms of  wrapping
numbers $(n_i^a,m_i^a)$ along the fundamental cycles $[a^i]$ and $[b^i]$ on
each $T^2$. Note also, that for  the specific orbifold   ${\bf Z_4\times Z_2}$
, the generators $\theta^2$ and $\omega$ are those of the ${\bf Z_2\times
Z_2}$ subgroup; these group elements transform each D-brane configuration into
itself, while $\theta$ and $\theta^3$ produce a {\it new image}  of the
original $D6$ brane configuration, and thus belong geometrically to a
different  open-string sector. We shall explicitly employ this geometric
feature  of the construction and obtain distinguished features of the spectrum
in different open string sectors.

\subsection{Homological Tadpole Cancellation and K-Theory Constraints}

The equation of motion for the Ramond-Ramond (R-R) field strength
$G_{8}=dC_{7}$ takes the form: \begin{equation} \label{tadd} {1\over
\kappa^2}\,
     d\star G_{8}=\mu_6\sum_a N_a\, \delta(\pi_a)+
                    \mu_6\sum_a N_a\, \delta(\pi_{a'})
        -4 \mu_6\,  \delta(\pi_{{\rm O}6}) \ ,
\end{equation}
 where $\delta(\pi_a)$ denotes the Poincar\'e dual three-form of $\pi_a$
cycles, $\pi_{a'}$ its orientifold image, $\kappa$ is the 10-dimensional
Planck constant and $\mu_6$ is the D6-brane tension.

Since the left hand side of eq. (\ref{tadd})  is exact, the R-R tadpole
cancellation condition reduces to  a simple condition on the homology classes
(see, \cite{BCLS} and references therein.):
\begin{equation} \label{tadpole} \sum_a  N_a\, (\pi_a + \pi'_a) -4 \pi_{O6}=0 \ .
\end{equation}

The above condition implies that the overall three-cycle wrapped by D-branes
and orientifold planes is trivial in homology. Again, for toroidal-type
compactifications   with D6-branes wrapping factorizable three-cycles
inherited from the six-torus,  the explicit expression (\ref{orcycle}), which
are specified by wrapping numbers $(n^i,m^i)$ along the fundamental cycles
$[a^i]$ and $[b^i]$, take a simpler  form written down in the following
sections.

K-Theory constraints  can be formulated \cite{Uranga} in terms of probe
D6-branes that wrap three-cycles of ${\rm O}6^i$ planes, and whose gauge
symmetry is $Sp(2k_i)$. The K-theory constraints imply that the massless
spectrum associated with the intersection of such probe D6-branes with the
D6-brane configurations of the model 
has an even number
of fundamental representations ${\bf 2k_i}$ of  $Sp(2k_i)$, and thus the
construction is free from discrete global anomalies \cite{Witten}. This
condition can again be expressed in terms of intersection numbers of cycles
associated with $\theta^i\omega^j\pi_{{\rm O}6}^T$ planes with the cycles
$\pi_a$ of the D6-brane configurations and can be  written  schematically in
the form:
\begin{equation}\label{Kth} (\pi_a + \pi'_a)\circ
(\theta^i\omega^j\pi_{{\rm O}6}^T)\in 2{\bf Z} \ .
\end{equation}

\subsection{Supersymmetry}

The supersymmetry condition for a three-cycle $\pi_a$ requires that it is a
special Lagrangian. Namely,  the restriction of the K\"ahler form $J$ of the
Calabi-Yau space  on the cycle vanishes, i.e., $J\vert_{\pi_a} =0$ and the
three-cycle is volume minimizing, i.e., the imaginary part of the three-form
$\Omega_3$ vanishes when restricted to the cycle, $ \Im(e^{i\varphi_a}\,
\Omega_3)\vert_{\pi_a} =0 $.
 The parameter $\varphi_a$ determines which ${ N}=1$ supersymmetry
is preserved by the branes. This  supersymmetry  condition also ensures that
  the Neveu-Schwarz-Neveu-Schwarz (NS-NS) tadpoles are cancelled as well.

For factorizable  three-cycles of toroidal compactifications these conditions
become geometric conditions:
\begin{equation} \label{susya} \phi^a_1+\phi^a_2+\phi^a_3=0\ {\rm mod}\ 2\pi,
\end{equation} where $\phi^a_i$ is the  angle with respect to the one-cycle on the i-th two-torus
 of the orientifold plane ${\rm O}6^1$ of Table
\ref{table1O6}. This condition can be rewritten in terms of $\tan\phi_i^a$'s
as:
\begin{equation} \label{susy} \sum_{i=1}^3\tan \phi_i^a=\prod_{i=1}^3\tan \phi_i^a\, , \ \
 \ \cos(\sum_{i=1}^3\phi_i^a)>0\, . \end{equation}  Again,
these condition can be expressed in terms of the three-cycle wrapping numbers
and toroidal complex structure moduli $U^{(i)} \equiv
\frac{R_2^{(i)}}{R_1^{(i)}}$.

\section{$SU(5)$ Grand Unified Model Constructions}\label{GUT}
In the following we shall  summarize the features of the spectrum and
couplings of the spectrum of the Grand Unified Models (GUT) for intersecting
D6-branes on ${\bf \bf Z_N\times Z_M}$ orientifolds.

The intersecting D6-branes on orientifolds  allow for the construction of GUT
models, based on the Georgi-Glashow $SU(5)$ gauge group.
 [Such supersymmetric Type IIA GUT constructions have a strong coupling limit, which is represented as
  a  lift  on a circle
 to M-theory compactified on singular seven  dimensional manifolds with $G_2$
holonomy  (see \cite{CSUIII}, and references therein).]

The key  feature in these constructions is  the appearance  of antisymmetric
(and/or symmetric) representations, i.e., ${\bf 10}$ (${\bf 15}$) of $SU(5)$,
which appear at the intersection of a  D6-brane with its orientifold image
(see Table \ref{tcs}). Therefore    ${\bf 10}$-plets, along with the
bi-fundamental representations $({\overline {\bf 5}}, \, { \bf N_b})$ at the
intersections of $U(5)$ branes with $U(N_b)$  (or $Sp(N_b)$)  branes,
constitute chiral fermion families of the Georgi-Glashow $SU(5)$ GUT model.
[Note that the gauge boson for the $U(1)$ factor of $U(5)$ is massive, and the
anomalies associated with this $U(1)$ are cancelled via the generalized
Green-Schwarz mechanism, ensured by the cancellation of the homological R-R
tadpole conditions (\ref{tadpole}) and the Chern-Simons terms in the expansion
of the Wess-Zumino D6-brane action. For details, as applied to the ${\bf Z_2
\times Z_2}$ orientifold, see the Appendix of \cite{CSUII}.]

For toroidal  orbifolds with D6-branes wrapping factorizable three-cycles,
inherited from the torus, there are three chiral superfields in the adjoint
representations on the world-volume of the D6-branes. In general, there are
additional chiral superfields in the adjoint representation, associated with
the intersections of the D6-brane configuration $a$ and its non-equivalent
orbifold images of the type $\theta^i \omega^j \, a$. The first set of fields
are moduli associated with the D6-brane splitting (or continuous Wilson
lines), and are a consequence of the fact that such cycles are not rigid. The
second set of fields are moduli associated with the brane recombination and
can also lead to further breaking of the $U(N_a)$ gauge symmetry. In the
effective theory, this geometric picture corresponds respectively to
 turning  on vacuum expectation values (VEVs)
 of  D-brane splitting  and D-brane recombination moduli fields, and it
   can  spontaneously  break  $SU(5)$
down to the Standard Model (SM) gauge group.  Since in the case of parallel
brane splitting all the  SM gauge group factors arise from D6-branes that wrap
parallel, homologically identical, cycles, this framework automatically
ensures that there is a gauge coupling unification.

[Within this framework one can in principle address  the  long standing
problem of doublet-triplet splitting, i.e., ensuring that after the  breaking
of $SU(5)$ the doublet of  ${\bf 5}_H$, the Higgs multiplet responsible for
the electroweak symmetry breaking,  remains light  while the triplet becomes
heavy. In the strong coupling limit, i.e., within M-theory compactified on
$G_2$ holonomy spaces \cite{WittenII}, this mechanism was addressed  via
discrete Wilson lines with different quantum numbers for the doublet and the
triplet fields. However, in the present context the Wilson lines, associated
with the D-brane splitting mechanism,  are continuous,
due to the  non-rigidity
of the three-cycles. This problem can be remedied by introducing rigid
three-cycles associated with  orbifold singularities, see \cite{BCMS}; however,
no explicit example of a chiral GUT model with discrete Wilson lines was found
there.]

The most important drawback of these constructions is the absence of Yukawa
couplings of the up-quark sector. Namely, the SM Higgs candidates  are
 in fundamental ($\overline{{\bf 5}}_H$ and ${\bf {5}}_H$) representations of
$U(5)$, and thus only Yukawa couplings of the type: $\overline{\bf 5}$~{\bf
10} $\overline{\bf 5}_H$ are present, while the couplings of the type ${\bf
10}~{\bf 10}~{\bf 5}_H$ are absent due to the $U(1)$ charge conservation. In
the strongly coupled limit of Type IIA theory, which corresponds to M-theory
compactified on singular  $G_2$ holonomy space, the absence of perturbative
Yukawa couplings to the up-quark families may be remedied by non-perturbative
effects, though see \cite{Berenstein}.\footnote{In the flipped $SU(5)$ context
\cite{NanopoulosI,NanopoulosII} the absence of such couplings corresponds to
those in the down-quark sector and thus their absence may be a somewhat less
severe problem.}

The  explicit supersymmetric GUT model  was first constructed on the ${\bf
Z_2\times Z_2}$ orientifold model \cite{CSUII}, and a systematic construction
of three-family models was given in \cite{CPS}. There are on the order of
twenty models with three families; however, they necessarily include in
addition to three copies of ${\bf 10}$-plets also three copies of {\bf
15}-plets,  i.e., the chiral supermultiplets in the symmetric representations
of $SU(5)$. These additional {\bf 15}-plets transform  under $SU(3)_C\times
SU(2)_Y \times U(1)_Y$ as: \begin{equation} {\bf 15}\to ({\bf 6},{\bf
1})(-\frac{2}{3}) + ({\bf 1},{\bf 3})(+1) + ({\bf 3},{\bf 2})(+\frac{1}{6}) \,
. \label{decomp}\end{equation} These multiplets therefore provide candidate
exotic SM fields, in particular,  ${\bf 3}$'s of $SU(2)_L$. Since  the ${\bf
15}$-plets can couple to $\bf{\bar 5}$ and ${\bar{\bf  5}}_H$,  after symmetry
breaking down to the Standard Model, triplets ${\bf 3}$ of $SU(2)_L$ could
couple to the  Standard Model Higgs fields and/or  leptons and could in
principle provide appropriate Majorana-type couplings for neutrino masses.
However, as described in the Introduction, for such a mechanism to be
effective one requires the $\bf 15$ to occur in a vector pair  of $({\bf
15}+{\overline{\bf 15}})$ which could become very massive, as well as the
needed couplings. [For further supersymmetric  constructions of such models,
see \cite{NanopoulosI}. For examples without chiral ${\bf 15}$-plets, which
are obtained  after the inclusion of Type IIA fluxes, leading to $AdS_4$
vacua, see \cite{NanopoulosII}.]

 Note, however, that for ${\bf Z_2\times Z_2}$ orientifold
models the orbifold images of the three-cycles wrapped by D6-branes are the
same as the original brane configuration,  and thus the spectrum associated
with the appearance of ${\bf 10}$ and ${\bf 15}$ arises from a single chiral
sector; therefore, in this case there is no possibility of generating from this
single sector  chiral ${\bf 10}$-plets  {\it and} vector-pairs of ${\bf 15}$
and ${\overline{\bf 15}}$-plets, associated with the genuine $N=2$
supersymmetric sector. Only in the case  where  such a mass spectrum can be
realized, could the $N=2$ vector pairs of ${\bf 15}$-plets  dynamically obtain
masses by the D-brane-splitting mechanism. On a specific two-torus where the
one-cycles of the brane configuration $a$, its orientifold image $a'$ and the
specific orientifold planes ${\rm O}6^i$ are parallel, the D-brane splitting
mechanism would be responsible for generating the mass for these pairs.

As we have emphasized earlier, for  other orbifolds, such as ${\bf Z_N\times
Z_M}$  the massless matter appears not only from the intersection of D6-branes
of the original configuration $a$ and its orientifold image $a'$, but also in
the sector associated with the orbifold images $\theta^i\omega^j\,  a$. In
particular, for the ${\bf Z_4\times Z_2}$  orbifold  for each  D6-brane
configuration $a$ there is the sector associated with its ${\bf Z_4}$ orbifold
image $\theta\, a$, and thus  some  sectors may result in chiral  and others
in non-chiral representations. It is this  feature of more general orbifolds
that we shall explore for the construction of the GUT Models which result in
the desirable vector-pair representations of the ${\bf 15}$-plets in one
sector and the chiral matter in the ${\bf 10}$ representation in another
sector.

\section{Model construction}\label{modelc}

We shall now apply the general tools described in
 Section   \ref{tools} for model construction to the case of the
 ${\bf Z_4\times Z_2}$ orientifold of Section   \ref{z2z4orien}.
 For this model the
form of the toroidal configuration and the ${\rm O}6$ planes are represented in
Figures \ref{MajorO1}-\ref{Majorth1}. The explicit  assignment of the ${\rm
O}6^i$ and $\theta {\rm O}6^i$ planes in terms of the wrapping numbers
$(n_i,m_i)$ for the basis one-cycles $[a_i]$ and $[b_i]$  is given in Tables
\ref{table1O6}-\ref{tableth1}.

\subsection{Global Consistency  Constraints}
From the general expressions for the global R-R consistency
conditions (\ref{tadpole}) we obtain the following conditions on the D-brane
wrapping numbers  (see also \cite{Honecker}):

\begin{eqnarray} \label{tad}
 &&    \sum_{a=1}^K   N_a  A \, n_3^a=32
  \nonumber \\
&&   \sum_{a=1}^K   N_a\, B\, (m_3^a+b\, n_3^a)=\,  -\frac{32}{4^b} \ ,
\end{eqnarray}
where
\begin{eqnarray} \label{def}
 A&\equiv&
n_1^a\, n_2^a\,-\, m_1^a\, m_2^a\, +\, n_1^a\, m_2^a\, +\, m_1^a\, n_2^a\, \nonumber\\
B&\equiv&  -\, n_1^a\, n_2^a\,+\, m_1^a\, m_2^a\, +\, n_1^a\, m_2^a\, +\,
m_1^a\, n_2^a \ ,
\end{eqnarray}
and $b=0,\frac{1}{2}$ for the third two-torus untilted and tilted,
respectively.  $A$ and $B$ are completely symmetric with respect to
the interchange of $(n_1^a,m_1^a)$ and $(n_2^a,m_2^a)$ wrapping numbers.

 In the calculation of homological R-R tadpole cancellation we have taken
into account for  each $a$ configuration with the wrapping numbers:
\begin{equation} a{\rm -configuration}:\ \  (n_i^a,m_i^a) \, , \ \ i=1,2,3\, ,
\end{equation}
its $\theta a$  image with the wrapping numbers:
\begin{equation} \theta a -{\rm image}: \ \ [(-m_1^a,n_1^2), (m_2^a,-n_2^a), (n_3^a,m_3^a)]\, , \end{equation}
and the orientifold images,  specified by the  wrapping numbers:
\begin{eqnarray}
a'-{\rm orientifold \ image}&:&\ \
(n_1^a,-m_1^a),(m_2^a,n_2^a),(n_3^a,-m_3^a-2b\, n_3^a)\, ,\\
\theta a'-{\rm orientifold\
image}&:&(-m_1^a,-n_1^a),(-n_2^a,m_2^a),(n_3^a,-m_3^a-2b n_3^a) .
\end{eqnarray} As for the action of the ${\bf Z_2\times Z_2}$ subgroup
elements, $\theta^2$, $\omega$ and $\theta^2\omega$, which render the D6-brane
configuration elements invariant, we appropriately normalized the intersection
numbers {\` a} la (\ref{orcycle}).

 Allowing for $4k_i$ branes wrapping the three-cycles of ${\rm
O}6^i$ planes (Table \ref{table1O6}) and their $\theta$ images $\theta {\rm
O}6^i$ (Table \ref{tableth1})  with the resulting gauge symmetry $Sp(2k_i)$,
the homological R-R tadpole cancellation condition (\ref{tad}) can be written
in the following way:
\begin{eqnarray} \label{tadp}
 &&    \sum_{a=1}^{K'}   N_a  A \, n_3^a=32-8\sum_{i=1,4}k_i
  \nonumber \\
&&    \sum_{a=1}^{K'}   N_a\, B\,(m_3^a+b\, n_3^a) =\, -\frac{32}{4^b}
+\frac{8}{4^b}\sum_{i=2,3}k_i \ ,
\end{eqnarray}
where now on the left-hand side the sum is only over the wrapping numbers of
the D6-brane configurations that are {\it not} parallel with the ${\rm O}6^i$
plane, and are associated with the $U(\frac{N_a}{2})$ gauge symmetry.

 The K-theory constraints  (\ref{Kth}) take the form:
\begin{eqnarray} \label{Kt}
 &&    \sum_{a=1}^K    N_a  A \, (m_3^a\, +\, b\, n_3^a)\in \frac{4 {\bf
 Z}}{4^b}
  \nonumber \\
&&    \sum_{a=1}^K   N_a\, B\, n_3^a \in 4{\bf Z} \ .
\end{eqnarray}
As discussed in  Section   \ref{tools}  these conditions ensure that the probe
branes wrapping cycles of ${\rm O}6^i$ branes (and their $\theta$ images),
which are associated with the  appearance of the $Sp(2k_i)$ gauge symmetry,
induce a massless spectrum at the intersections with  the $U(N_a/2)$ D6-branes
that have an {\it even} number of chiral superfields in the fundamental ${\bf
2k_i}$ - representation of $Sp(2k_i)$, and it is thus free of discrete global
gauge anomalies \cite{Witten}.

\subsection{Supersymmetry Constraints}
For the supersymmetry conditions (\ref{susy}), the expressions for $\tan \phi_i^a$
of the  angles $\phi_i^a$ with respect to the ${\rm O}6^1$ plane take the form
\cite{Honecker}:
\begin{equation}
\tan\phi_1^a=\frac{m_1^a}{n_1^a}\, , \ \
\tan\phi_2^a=\frac{(m_2^a-n_2^a)}{(n_2^a+m_2^a)}\, , \ \
\tan\phi_3^a=-\frac{(m_3^a+b\, n_2^a)U^{(3)}}{n_3^a}\, , \end{equation} where
$U^{(3)}\equiv {{R_2^{(3)}}\over {R_1^{(3)}}}$ is the complex structure
modulus for the third two-torus. The supersymmetry conditions (\ref{susy}) in
turn take the form
\begin{eqnarray}\label{susy0}
\frac{m_1^a}{n_1^a}+\frac{(m_2^a-n_2^a)}{(n_2^a+m_2^a)}+\frac{(m_3^a+bn_3^a)\,
U^{(3)}}{n_3^a}&=&\frac{m_1^a}{n_1^a}\frac{(m_2^a-n_2^a)}{(n_2^a+m_2^a)}\frac{(m_3^a+bn_3^a)\,
U^{(3)}}{n_3^a}\, ,\nonumber\\
-An_3^a+B(m_3^a+b\, n_3^a)U^{(3)}&\le& 0\, .
\end{eqnarray}
We  can  solve the above conditions  for  $U^{(3)}$:
\begin{equation}\label{U}
U^{(3)}=-\frac{n_3^a\, B}{(m_3^a\,+\, b\, n_3^a)A}>0\, , \ \
\frac{n_3^a}{A}(A^2+B^2)\ge 0\, .
\end{equation}
These conditions provide strong constraints on the allowed wrapping
numbers $(n_i^a,m_i^a)$ of different D6-brane configurations, since (\ref{U})
should be  satisfied for each such configuration.

 \subsection{ Spectrum}

 The gauge symmetry of $N_a$-D6 branes, with the
general wrapping numbers $(n_i^a,m_i^a)$ is $U(N_a/2)$.  For $N_a$ D6-branes
positioned at any of the ${\rm O}6$ planes in Table \ref{table1O6} (and
depicted in Figure  \ref{MajorO1})), the additional orientifold projection
introduces an  $Sp(N_a/2)$ gauge group in the open string sector on the
D-brane, i.e., in this case a multiple of 4 branes is needed.   The spectrum
on the $U(\frac{N_a}{2})$ D6-brane consists of 3-fields in the adjoint
representation, corresponding to the D-brane splitting moduli  as well as
\cite{Honecker}
\begin{equation}
I_{a\theta a}'= [(n_1^a)^2+(m_1^a)^2][(n_a^2)^2+(m_2^a)^2]  \end{equation}
D-brane recombination moduli, also in the $N=2$ sector of the adjoint
representation. In the $Sp(2k_i)$ sector, the multiplicity of the D-brane
splitting and recombination moduli is $3$ and $2$, respectively, and they are
in the symmetric representation of $Sp(2k_i)$ symmetry group \cite{Honecker}.

 In the following we shall focus on the symmetric and
anti-symmetric representations of the D-brane configurations, associated with
the GUT $SU(5)$ symmetry group. As discussed in  Section   \ref{tools}, we shall
split the calculation into sectors associated with $a$-unitary brane
configuration and its $Z_4$ orbifold image $\theta a$. Employing the
expression for the intersection numbers (Table \ref{tcs}, and eqs.
(\ref{orcycle})) we obtain the following expressions for the intersection
numbers:
\begin{eqnarray} I_{a\, a'}&=&4n_1^a\, m_1^a\, [(n_2^a)^2-(m_2^a)^2]\, n_3^a\,
(m_3^a+b\, n_3^a)\, , \nonumber
\\I_{(\theta a)\, a'}&=&4\, [(n_1^a)^2-(m_1^a)^2]\,n_2^a\, m_2^a\,  n_3^a\,
(m_3^a+b\, n_3^a)\, .
\end{eqnarray}
Note that $I_{a\, (\theta a')}=I_{(\theta a)\, a'}$.

The intersection of the $a$-D6-brane configuration with the  orientifold
planes can be split into the one with intersections with   ${\rm
O}6_{tot}\equiv \sum_{i=1}^4 {\rm O}6^i$ planes  (see Table \ref{table1O6} for
the wrapping numbers of ${\rm O}6^i$) and the second one, for intersections
with $\theta{\rm O}6_{tot}\equiv \sum_{i=1}^4\theta{\rm O}6^i$ (see Table
\ref{tableth1} for the wrapping numbers of $\theta{\rm O}6^i$). Due to the
symmetry of the configurations it turns out that these intersection numbers
are  the same for both sectors, and they take the form:
\begin{equation} I_{a\, {\rm O}6_{tot}}=I_{a\, \theta{\rm O}6_{tot}}=4[A\, (m_3^a+b\,n_3^a)+B\, b\, n_3^a)]\, .
\end{equation}
Due to the symmetry of the construction the intersection number of the $\theta
a$ configuration with the  ${\rm O}6_{tot}$  and $\theta{\rm O}6_{tot}$ sector
has the same intersection number as above, i.e., \begin{equation}
\label{orienint}I_{(\theta a)\, {\rm O}6_{tot}}=I_{a\, {\rm O}6_{tot}}\,
.\end{equation}

The multiplicity of the symmetric and anti-symmetric representations in $a$
and $\theta a$ sectors  is determined by the following expressions:
\begin{eqnarray}\label{asi}
I_a^{symm, antisymm}&=&{\frac{1}{2}}\left(I_{a\, a'}\pm \frac{1}{2}I_{a\, {\rm
O}6_{tot}}\right)\, , \nonumber\\
I_{\theta a}^{symm, antisymm}&=&{\frac{1}{2}}\left(I_{(\theta a)\, a'}\pm
\frac{1}{2}I_{(\theta a)\, {\rm O}6_{tot}}\right)\, .
\end{eqnarray}
In the calculation of the intersection numbers (\ref{asi}) we have accounted
for the multiplicity of the equivalent configurations associated with the
${\bf Z_2\times Z_2}$-part of the  orbifold action, i.e., those associated
with  ${\bf Z_2\times Z_2}$ group elements: $\theta^2$, $\omega$ and
$\theta^2\omega$.

An important observation is in order: since  there are two separate sectors,
associated with the appearance of symmetric and anti-symmetric representations,
there is now a possibility that in one sector the representations are for
example ${\bf 15}$-plets  and in another  sector, ${\overline{\bf 15}}$-plets
(and analogously for the anti-symmetric representations). However, note
that such  ${\bf 15}$-plets and ${\overline{15}}$ arise from the $N=1$ sector and
are thus chiral in nature.

We require there are only chiral ${\bf 10}$ representations, and no net chiral
${\bf 15}$-plets.
 In addition, we shall require
that there is a genuine $N=2$ sector associated with the vector pairs of ${\bf
15}$  (and ${\bf 10}$).  The necessary conditions to ensure  these constraints
are: say, $I_{(\theta a)\, a'}=0$ and  $I_{a\, a'}=I_{a\, {\rm O}6_{tot}}\ne
0$.  Namely, the first condition is a necessary condition to have a genuine
$N=2$ sector, and the second condition  then automatically ensures that there
are no net chiral ${\bf 15}$-plets (see eq. (\ref{asi}))\footnote{Note that in
the case of ${\bf Z_2\times Z_2}$ orientifold one has only the sector
associated with $a$ and $a'$ configuration. Therefore the necessary condition
to have vector pairs requires $I_{aa'}=0$  which automatically implies the
same number of chiral ${\bf 10}$'s and ${\overline{\bf 15}}$'s (or
${\overline{\bf 10}}$'s and ${\bf 15}$'s).}. To address quantitatively the
appearance of vector pairs in the $N=2$ sector, we have to focus on sectors
that involve the two-torus where both the one-cycle for  $a$ and $\theta a'$
configurations are parallel as well as that of specific $\theta {\rm
O}6^i$-planes. In  this subsector there are consequently {\it no}
intersections  associated on the specific two-torus, and the spectrum is that
of $N=2$  vector pairs, which can be determined in terms of the intersection
numbers on the remaining four-torus as:
\begin{equation}
I_{a;N=2}^{symm, antisymm}=I_{a\,\theta a'}'\pm \sum_{i} I_{a\, \theta{\rm
O}6^{i}}'\, ,
\end{equation}
where $\prime$ refers to the intersection numbers in the remaining four-torus
and the summation is only over  intersections with $\theta{\rm O}6^{i}$-planes
which are parallel with the one-cycle of $a$ configuration in the specific
two-torus.

The conditions for the number of bi-fundamental representations associated
with $a$ and $b$ branes take the following expression  (see also
\cite{Honecker}):
\begin{eqnarray}
&&I_{a,b}+I_{a,\theta b}=(-n_3^a\,m_3^b+m_3^a\, n_3^b) \nonumber\\
&&\times [(n_1^am_1^b-m_1^an_1^b)(-n_2^a\, m_2^b+m_2^a\,n_2^b)+
 (n_1^an_1^b+m_1^am_1^b)(n_2^a\, n_2^b+m_2^a\,m_2^b)]  \, , \nonumber\\
&&I_{a,b'}+I_{a,\theta b'}=(+2b\, n_3^a n_3^b+n_3^a\,m_3^b+m_3^a\,
n_3^b) \nonumber\\
&&\times [(n_1^am_1^b+m_1^an_1^b)(n_2^a\, n_2^b-m_2^a\,m_2^b)+
 (n_1^an_1^b-m_1^am_1^b)(n_2^a\, m_2^b+m_2^a\,n_2^b)] \, .\end{eqnarray}
We would like reiterate that  we have chosen a convention that the left-handed
chiral superfields in representations, according to Table \ref{tcs},
correspond to the negative values of the above intersection numbers. (For the
positive values of these intersection numbers the left-handed chiral field
representations correspond to the charge-conjugated ones.)

At this point we are equipped with all the tools to construct a specific
model, with relevant interesting implications for neutrino masses.

\section{Explicit four-family GUT Model}\label{explm}

A specific, globally consistent supersymmetric  model with the wrapping
numbers of D6-branes and their intersection numbers (expressions given in the
previous Section   \ref{modelc}) is depicted in Table \ref{4GUT}.  The
explicit chiral and non-chiral spectrum is presented in Table \ref{spectrum}.
For this model   the homological (\ref{tadp})  and K-theory (\ref{Kt}) tadpole
constraints (with respective contributions  $-\frac{3}{2}\times (10+2)$ and
$1\times (10+2)$) are satisfied, and the supersymmetry conditions
(\ref{susy0}) are satisfied for the value of the complex structure modulus
$U^{(3)}=\frac{2}{3}$.

\begin{table}
[htb] \footnotesize
\renewcommand{\arraystretch}{1.0}
\caption{D6-brane configurations and intersection numbers for globally
consistent  four family Grand Unified model.   } \label{4GUT}
\begin{center}
\begin{tabular}{|c||c|c||c|c|c|c|c|}
\hline
     & \multicolumn{7}{c|}{$U(5)\times U(1) \times
    \times Sp(4)_2$}\\
\hline \hline \rm{stack} & $N$ & $(n^1,m^1)\times (n^2,m^2)\times
(n^3,m^3)$ & $n_{\Ysymm}$& $n_{\Yasymm}$ &  $b$  & $b'$&1  \\
\hline \hline
    $a$&  10& $(1,2)\times (1,0)\times (1,-1)$ & (5+1)$\times${\rm pairs} & 4 +1$\times${\rm pair}  & 2  & -2& 1  \\
    $b$&  2& $(1,0)\times (0,1)\times (1,-2)$ & -2 & 0  &  &  &1\\

\hline \hline
    1&   8 & $(1,0)\times (1,-1)\times (0,1)$ &  & & & &\\
\hline
\end{tabular}
\end{center}
\end{table}

\begin{table}
[htb] \footnotesize
\renewcommand{\arraystretch}{1.25}
\begin{center}
\begin{tabular}{|c||c||c|}
\hline
 Sector & $U(5)\times U(1)\times Sp(4)_2$& Fields
\\
\hline
 $a  a$ & $(3 \, +5)\times  (
 {\bf 25},0,{\bf 1})\, $& D-brane-splitting + recombination moduli \\
 $b  b$ & $(3 \, +1) \times  (
 {\bf 1},0,{\bf 1})\, $& D-brane-splitting + recombination moduli\\
 $c  c$ & $(3 \, +2) \times  (
 {\bf 1},0,{\bf 6})\, $& D-brane-splitting + recombination moduli\\ $a  a'$ & $3 \times  (
 {\bf 10},0,{\bf 1})\, +\, 1\times ({\bf 15},0,{\bf 1})$& chiral ${\bf 10}$-fermion
 families+
 chiral-${\bf 15}$\\

$\theta a a'$ & $1 \times ({\bf 10},0,{\bf 1})$ +$1\times ({\overline{\bf
15}},0,{\bf 1})$&chiral ${\bf 10}$-fermion family+ chiral-$\overline{{\bf 15}}$\\
          & $1 \times ({\bf 10}+\overline{{\bf 10}},0,{\bf 1})$+
          $5 \times ({\bf 15}+\overline{{\bf 15}},0,{\bf 1})$& non-chiral
          pairs of ${\bf 10}$+ ${\bf 15}$
          \\
       $b b'$
          & $1 \times ({{\bf 1}},-2,{\bf 1})$& ``hidden sector'' chiral matter
           \\
$\theta b b'$          & $1 \times ({\bf 1},-2,{\bf 1})$& ``hidden sector'' chiral matter\\
$a b$
          & $2 \times ({\overline{\bf 5}},1,{\bf 1})$& chiral ${\bf \bar 5}$-fermion
          families
           \\
$a b'$          & $2 \times ({\bf 5},1,{\bf 1})$& chiral SM up-Higgs\\
$a c$ & $1\times (\overline{{\bf 5}},0,{\bf 4})$& chiral ${\bf\bar 5}$-fermion
families +
SM down-Higgs\\
$b c$ & $1\times ({\bf 1},-1,{\bf 4})$ & ``hidden sector'' chiral matter\\
\hline
\end{tabular}
\end{center}
\caption{\small The chiral spectrum  and non-chiral spectrum, as obtained from
the information for the configuration and the intersection numbers
 listed in Table \ref{4GUT}.  \label{spectrum}}
\end{table}

The gauge symmetry of the model is $U(5)\times U(1)\times Sp(4)_2$, where
$Sp(4)_2$ is associated with the $\Omega R\omega$ action, i.e.,
${\rm
O}6^2$-plane in Table \ref{table1O6}.  One can satisfy the homological and
K-theory tadpoles also by replacing $Sp(4)_2$ with $Sp(2)_2\times Sp(2)_3$,
where these gauge group factors arise from the D6-branes on ${\rm O}6^2$ and
${\rm O}6^3$ planes, respectively.

In addition to the four-family chiral spectrum $4\times ({\bf 10}+
{\overline{\bf 5}})$, the model possesses two pairs of the Standard model
Higgs candidates $2\times ({\bf 5}+\overline{\bf 5})$. The $N=2$ non-chiral
sector consists of
 5 vector pairs of (${\bf 15}+{\overline{\bf 15}}$) and 1 vector pair of (${\bf 10}+{\overline{\bf 10}}$).
These vector pairs can obtain a mass due to the parallel D-brane splitting
in
the second two-torus. There is an additional vector pair of (${\bf
15}+{\overline{\bf 15}}$); however,
 its origin is chiral, i.e., it arises from the $N=1$ sector, and thus it cannot obtain a mass from
 parallel D-brane splitting.

In  general, one would expect that there are  non-zero  Yukawa couplings of
both ${\bf 10}$'s as well as  ${\bf 15}$'s to  bi-linears of ${\bf \bar 5}$'s,
which play a role of down-sector Standard Model Higgs fields and/or lepton
doublets. In principle there could also be  Yukawa couplings of $\overline{\bf
10}$'s as well as $\overline{\bf 15}$'s  to bi-linears of $\bf 5$'s, which are
the up-sector Standard Model Higgs candidates. (For the   full conformal field
theory calculation of such couplings see \cite{CP}, and for a detailed
analysis of the  classical part of the Yukawa couplings, see \cite{IbaY}.)
However, for  the specific construction the only surviving Yukawa couplings
are those of bi-linears of $({\bf \bar 5}, 0, {\bf 4} )$'s to $({\bf 10}, 0,
{\bf 1})$'s, two components of $({\bf \bar 5}, 0, {\bf 4})$'s playing a role
of the Standard Model (SM) down-Higgs, and two components corresponding to two
fermion families.\footnote{There are also Yukawa couplings of $({\bf
5},1,{\bf 1})\times ({\bar {\bf 5}}, 0, {\bf 4})\times ({\bf 1}, -1,{\bf
4})$.} Unfortunately, due to the gauge invariance constraints, in this
specific construction  the Yukawa couplings to ${\bf 15}$'s, $\overline{\bf
15}$'s (as well as $\overline{\bf 10}$'s)  are absent. Had the couplings of
${\bf 15}$ and $\overline{\bf 15}$ vector pairs to $\bf 5$ and
 ${\bar{\bf 5}}$ multiplets been present, they would have played
an important role for generating Majorana type neutrino masses. Note, however,
that in principle in other related constructions there does not seem to be any
obstruction for  such couplings to exist.

\subsection{Flipped SU(5) GUT Interpretation}\label{flip}

We would also like to  point out that the above construction can be
interpreted as a   flipped $SU(5)$ construction.  A linear combination of
$U(1)_5$ of $U(5)$, $U(1)\equiv U(1)_1$ and the Abelian subgroup $U(1)_4$ of
$Sp(4)$ (which can be obtained after the D-brane splitting
mechanism\footnote{For such a D6-brane splitting analysis as well as a
complementary field theoretical Higgs mechanism, see \cite{CLLL}.})  provide
an adequate $U(1)_X$  of the flipped $SU(5)$ construction:
\begin{equation} Q_X=\frac{1}{4}[Q_5-5(Q_1\pm Q_4)]\, .
\end{equation} (For a brief summary of features of the  flipped $SU(5)$ see, e.g., \cite{NanopoulosI}.)
For the specific $Q_X$ charges of the model and the spectrum interpretation,
see Table \ref{flippeds}.

This specific combination of $Q_X$  turns out to be non-anomalous, i.e. the
gauge boson for $U(1)_X$ is massless. This result could be suspected from the
absence of field theoretical triangular anomalies associated with the $U(1)_X$
gauge field, and therefore  the generalized Green-Schwarz contributions to
these anomaly cancellations are absent \footnote{For a detailed analysis of
the cancellation of gauge and gravitational anomalies for the ${\bf Z_2\times
Z_2}$, see the Appendix of \cite{CSUII}. A generalization to other
orientifolds is straightforward, but it involves a careful bookkeeping of the
orbifold images of  D6-brane configurations.}. In the following we show that
the Chern-Simons term, which plays a role in the generalized Green-Schwarz
mechanism and is  responsible for the mass of the $U(1)_X$ gauge boson, is
indeed absent.
\begin{table}
[htb] \footnotesize
\renewcommand{\arraystretch}{1.35}
\begin{center}
\begin{tabular}{|c||c|c||c|}
\hline
 Sector & Flipped $U(5)\times U(1)\times Sp(4)_2$& $U(1)_X$&Fields
\\
\hline
 $a  a'$ & $3 \times  (
 {\bf 10},0,{\bf 1})\, +\, 1\times ({\bf 15},0,{\bf 1})$& $\frac{1}{2}$& chiral ${\bf 10}$-fermion
 families+
 ${\bf 15}$\\

$\theta a a'$ & $1 \times ({\bf 10},0,{\bf 1})$ +$1\times ({\overline{\bf
15}},0,{\bf 1})$& $\frac{1}{2}$ + $(-\frac{1}{2})$&chiral ${\bf 10}$-fermion family+ $\overline{{\bf 15}}$\\
          & $1 \times ({\bf 10}+\overline{{\bf 10}},0,{\bf 1})$+
          $5 \times ({\bf 15}+\overline{{\bf 15}},0,{\bf 1})$&$\frac{1}{2}+\, (-\frac{1}{2})$& non-chiral
           ${\bf 10}$-GUT Higgs pair + ${\bf 15}$-pairs
          \\
         $b b'$
          & $1 \times ({\bf 1},-2,{\bf 1})$& $\frac{5}{2}$&chiral charged lepton\\
$\theta b b'$          & $1 \times ({\bf 1},-2,{\bf 1})$& $\frac{5}{2}$&chiral charged lepton\\
$a b$
          & $2 \times ({\overline{\bf 5}},1,{\bf 1})$& $-\frac{3}{2}$&chiral ${\bf \bar 5}$-fermion families\\
$a b'$          & $2 \times ({\bf 5},1,{\bf 1})$& $-1$&chiral SM up-Higgs\\
$a c$ & $1\times (\overline{{\bf 5}},0,{\bf 4})$&$2\times(-\frac{3}{2}+1)$&
chiral ${\bf\bar 5}$-fermion families +
SM down-Higgs\\
$b c$ & $1\times ({\bf 1},-1,{\bf 4})$ &$2\times(\frac{5}{2}+0)$& chiral charged leptons +exotics\\
\hline
\end{tabular}
\end{center}
\caption{\small The charge assignments and the matter spectrum interpretation
for a  flipped $SU(5)\times U(1)_X$ GUT. $U(1)_X$ is shown to be
non-anomalous.  \label{flippeds}}
\end{table}

The specific  Chern-Simons term, responsible for the mass of the $U(1)_a$
gauge field for the D6-brane configuration $a$, arises in the expansion of the
D6-brane Wess-Zumino action. (For details and an application to ${\bf
Z_2\times Z_2}$-orientifold see the Appendix of \cite{CSUII}.) It has the
following form:
\begin{equation}  2N_a\, (p^a_I+p^{\theta a}_I) \, \int_{\bf R^{1,3}}\, B^I\wedge
F_a\, . \label{cs}\end{equation}   In the above expression, the factor of two
accounts for the same contribution from the orientifold images, $F_a$ is the
$U(1)_a$ gauge field strength, $B^I$'s are the two-form fields (dual to the
axion fields $\Phi_I$ of toroidal moduli), and ($p^a_I,\ p^{\theta \, a}_I)$
are respective wrapping numbers of the D6-brane configuration $a$ and its
$\theta \, a$ image
 along the the $I$-th three-cycle of the  lattice $\Lambda^I$:
\begin{equation}
\Lambda^I=\{-[b_1^o]\times [b_2^o]\times [b_3^o],\ [b_1^o]\times [a_2^o]\times
[a_3^o],\ [a_1^o]\times [b_2^o]\times [a_3^o],\ [a_1^o]\times [a_2^o]\times
[b_3^o]\}\,
\end{equation}
where $([a_i^o],[b_i^o])$ are the one-cycles, parallel with and perpendicular
to the  ${\rm O6}^1$-plane, respectively.  Note, $\Lambda^I$ is dual to the
lattice $\Sigma_I$:
\begin{equation}
\Sigma_I=\{[a_1^o]\times [a_2^o]\times [a_3^o],\ [a_1^o]\times [b_2^o]\times
[b_3^o],\ [b_1^o]\times [a_2^o]\times [b_3^o],\ [b_1^o]\times [b_2^o]\times
[a_3^o]\}\,,
\end{equation}
with the property that $\Lambda^I \circ \Sigma_J=\delta^I_J$.

\begin{table}
[htb] \footnotesize
\renewcommand{\arraystretch}{1.0}
\caption{Wrapping numbers $p^a_I$ and $p^{\theta\,  a}_I$ along the
three-cycles of the dual lattice $\Lambda^I$. Again, $b=0,\ \frac{1}{2}$ for
the untilted and tilted third two-torus, respectively. } \label{ps}
\begin{center}
\begin{tabular}{|c||c|c|}
\hline $I$ & $p^a_I$ & $p^{\theta\, a}_I$\\
\hline\hline
    $1$& $m_1^a (n_2^a-m_2^a)(m_3^a+bn_3^a)$ & $n_1^a (n_2^a+m_2^a)(m_3^a+bn_3^a)$\\
\hline
    $2$& $m_1^a (n_2^a+m_2^a)n_3^a$ & $-n_1^a (n_2^a-m_2^a)n_3^a$\\
\hline
    $3$& $-n_1^a (n_2^a-m_2^a)n_3^a$ & $m_1^a (n_2^a+m_2^a)n_3^a$\\
\hline
    $4$& $n_1^a (n_2^a+m_2^a)(m_3^a+bn_3^a)$ & $m_1^a (n_2^a-m_2^a)(m_3^a+bn_3^a)$\\
\hline
\end{tabular}
\end{center}
\end{table}
For a  configuration with wrapping numbers $(n^a_i,m^a_i)$ (with respect to
the basis one cycles $([a_i],[b_i])$ of the original six-torus),  the values
of $(p^a_I,p^{\theta\, a}_I)$ are listed in  Table \ref{ps}.  The contribution
of the  Chern-Simons terms (\ref{cs}) to  the $U(1)_X$ field strength involves
the following  linear combination of the coefficients $p_I$: \begin{equation}
 5 (p^a_I+p^{\theta a}_I) -
5(p^b_I+p^{\theta\, b}_I) \, , \end{equation}
 where $a$ and $b$  refer to the
respective configurations for $U(5)$ ($N_a=10$) and $U(1)$ ($N_b=2$), and
whose wrapping numbers are given in Table \ref{4GUT}. Note also, that it is
the linear combination of $\frac{1}{4}(Q_5-5Q_1)$ charges that contributes to
$U(1)_X$, and that $U(1)_4$, since it arises from the non-Abelian gauge
symmetry $Sp(4)$, is automatically non-anomalous.  It is now straightforward
to show that these linear  combinations  are indeed zero for {\it all four}
I's.

While $U(1)_X$ is a suitable anomaly free candidate for the flipped
$SU(5)\times U(1)_X$, the model suffers from a number  of phenomenological
problems. There is also the absence of Yukawa couplings of the GUT Higgs
candidates ${\overline{\bf 10}}$ to the SM Higgs candidates ${\bf 5}$'s, and
thus the model  does not address a part of doublet-triplet splitting problem,
and, of course, the model also suffers from the absence of the down-sector
Yukawa couplings (just as the standard $SU(5)$ GUT's  in this framework do not
have perturbative top-sector Yukawa couplings). Nevertheless, the
constructions of that type provide a net number of chiral ${\bf 10}$-plets
(and no net number of chiral ${\bf 15}$-plets) as well as potential flipped
$SU(5)$ GUT Higgs candidates, as $({\bf 10}+{\overline{\bf 10}})$ vector
pairs.

In the conclusion of this Section we would like to emphasize that  although at the
coupling level we are faced with specific obstacles, the explicit construction
presented in this paper provides us with a geometric approach to identify a
desirable massless spectrum of  GUT constructions. We would also like to
emphasize that the geometric interpretation of the origin of the spectrum in
our case  allows for the clear identification of genuine $N=2$ vector pairs of
${\bf 15}$-plets as well as those that arise from the $N=1$ sector. Therefore,
we are able to determine which pairs can obtain mass after D-brane splitting
and which are protected due to their  chiral origin. At the coupling level we
also have explicit techniques to calculate  Yukawa couplings, although  zero
values of such couplings are typically determined already at the level of
gauge invariance.

 \section{Conclusions }\label{conc}

 In this paper we have provided detailed technical tools for the construction of $SU(5)$ grand-unified models
 (GUT's) with intersecting D6-branes on orbifolds different from $T^6/({\bf Z_2\times
 Z_2})$ orientifolds. Specifically, we chose the $T^6/({\bf Z_4\times Z_2})$
 orientifold and the three-cycles wrapped by D6-branes that are inherited from
 the original six-torus $T^6$. In particular, we highlighted the new features
 of the spectrum, that allows in addition to the chiral sector of ${\bf 10}$'s
 and ${\bf \bar 5}$'s also the appearance of the vector pairs of ${\bf 15}$'s
 and ${\bf 10}$'s. In the genuine $N=2$ sector such vector pairs can obtain a
 mass due to a parallel D6-brane splitting in a specific two-torus.
We have constructed such a globally consistent, supersymmetric model with
$U(5)\times U(1)\times Sp(4)$ gauge symmetry,  four-families of $({\bf
10}+{\bar{\bf 5}})$'s and two pairs of the Standard Model Higgs candidates.
 In the
 case that ${\bf 15}$'s  (and $\overline{\bf 15}$) could couple  to ${\bf {\bar 5}}$ (and $\bf 5$) bi-linears,
such Yukawa couplings would provide, after symmetry breaking (via parallel
D-brane splitting) down to the Standard Model (SM),   the relevant couplings
to generate small Majorana-type masses for left-handed neutrinos.
Unfortunately, such couplings are not present in the concrete construction,
although we do not see any obstruction in principle to having such couplings in
related constructions.

We have also pointed out that this construction can  have an interpretation of
the flipped $SU(5)\times U(1)_X$ GUT model, where we have shown that the $U(1)_X$
gauge boson  remains massless. For this interpretation the $({\bf
10}+{\overline{\bf 10}})$-vector pairs can play the role of the GUT Higgs
candidates, while there is only a net number of chiral $({\bf 10}+{\bf \bar
5})$'s, as family candidates, i.e. there are no net chiral ${\bf 15}$'s. The
concrete model has two  additional singlets which could play a role of
right-handed neutrinos.  There are some phenomenological problems at the
Yukawa coupling level. Nevertheless we expect that related constructions may
well produce more realistic flipped $SU(5)$ GUT models with interesting
phenomenological implications.

With an aim to construct related models that pass the tests not only at the
spectrum but also at the coupling level,
 we plan to turn to constructions of models on other orientifolds as
 well as three-cycles associated with the orbifold singularities.

\section*{Acknowledgments}

We would like to thank  Ralph Blumenhagen, Elias Kiritsis, Michael Schulz and
Robert Richter  for useful discussions. We are grateful to the Galileo Galilei
Institute for Theoretical Physics and Aspen Center for Physics for hospitality
during the course of the work. The research was supported in part by the
National
 Science Foundation under Grant No.~INT02-03585 (MC),
 Department of Energy Grant
 DOE-EY-76-02-3071 (MC,PL)
and  the Fay R. and Eugene L. Langberg Chair  (MC).

\end{document}